# Legal Architecture and Design for Gulf Cooperation Council Economic Integration

Bashar H. Malkawi

TABLE OF CONTENTS

## ABSTRACT

The Cooperation Council for the Arab States of the Gulf (GCC) is generally regarded as a success story for economic integration in Arab countries. The idea of regional integration gained ground by signing the GCC Charter. It envisioned a closer economic relationship between member states. Although economic integration among GCC member states is an ambitious step in the right direction, there are gaps and challenges ahead. The best way to address the gaps and challenges that exist in formulating integration processes in the GCC is to start with a clear set of rules and put the necessary mechanisms in place. Integration attempts must also exhibit a high level of commitment in order to deflect dynamics of disintegration that have all too often frustrated meaningful integration in Arab countries. If the GCC can address these issues, it could become an economic powerhouse within Arab countries and even Asia.

**I. Introduction**

The current era is characterized by the proliferation of regional trade agreements around the world.[1] In light of the slow progress made to conclude the Doha Round of the World Trade Organization (WTO), an avalanche of bilateral and regional free trade agreements will fill in the vacuum. The legacy of the failure of multilateralism is a renewed global push toward bilateralism.

Arab countries have embarked upon ambitious continental integration efforts designed to fulfill their developmental goals.[2] The principles surrounding Arab economics, their economic integration focus, are the same as for any regional integration: combining the resources of constituent members in an effort to achieve economies of scale, comparative advantages, and development.[3] Arab countries have had several sub-continental regional arrangements for years.[4] However, the Cooperation Council for the Arab States of the Gulf (GCC) is generally regarded as a success insofar as its incarnations as a free trade area, a customs union, and a common market are concerned. The outcome of the GCC is still an unfolding drama of integration.

---

[1] Looking at regional integration, one can immediately see the upward pattern of the trend. Between 1978 and 1991, the number of regional trade agreements (RTAs) remained nearly static. Since the beginning of the 1990s, the trend was reversed and one could observe a constant dramatic increase in the number of RTAs that are being formed. From 42 RTAs notified to the General Agreement on Tariffs and Trade (GATT) according to Article 7(a) of the GATT in 1991, the number increased by 107% to 87 Agreements in 1998. See Matthew W. Barrier, Regionalization: The Choice of a New Millennium, 9 Currents: International Trade Law Journal 25, 27 (2000). According to the World Trade Organization (WTO), there are currently 170 RTAs in force. The WTO expects the total number of RTAs to rise to nearly 320 by the end of 2009.
[2] Arab Countries are: Algeria, Bahrain, Comoros, Djibouti, Egypt, Iraq, Jordan, Kuwait, Lebanon, Libya, Mauritania, Morocco, Oman, Palestinian Autonomous Territories, Qatar, Saudi Arabia, Somalia, Sudan, Syria, Tunisia, United Arab Emirates, and Yemen.
[3] The theory of comparative advantage suggests that countries will trade goods which they are most efficient in producing. Also, trading blocks create trade opportunities (trade creating) though admittedly with some trade diversion. Trade creation suggests that there will be more opportunities for efficient operations to excel, in large part because of the rearrangement of productive tasks among countries with different comparative advantages - tasks that lead to freely tradable goods within economic integration, which in turn results in the development of more efficient operations due to the increased competition. The increased competition is accompanied by, however, increased market size and, in theory, an increase in economic activity. See Raj Bhala, International Trade Law: Theory and Practice 635-650 (2d ed. 2001).

The GCC was established in May of 1981. The GCC consists of six member states: (1) United Arab Emirates; (2) Bahrain; (3) Saudi Arabia; (4) Oman; (5) Kuwait; and (6) Qatar.[5] While there are many elements which led to the establishment of the GCC, chief among them was to foster economic integration between members, increase their bargaining power in international relations and, through collective security, to guard against any threat from neighboring states. More specifically, the GCC was formed against the backdrop of the Islamic revolution in Iran and the Iraq-Iran war. In those state of affairs, GCC credibility depended on the ability of its member states to consolidate and build upon the common foreign and security concerns they face. In addition, the differences among the member states of GCC are negligible as they share history, political structures, languages, and cultural habits.[6] Structurally, the GCC is helped along by the fact that it has a manageable number of states and a high level of development. Thus, GCC members share an already existent common identity and cohesion.

---

[4] The fact that the experiences Arab countries have had with integration have not been good or successful ones. Several efforts have been made to jump start the integration process, yet none have successfully moved integration toward completion of its mandate. Several treaties were signed to accomplish this purpose: (1) the Joint Defense and Economic Cooperation Treaty among Member States of the League of Arab States in 1950, (2) the Convention for Facilitating Trade and Regulating Trade Transit in 1953, and (3) the Arab Economic Unity Agreement in 1957. In addition, some attempts for Arab Common Market took place in 1964. The most recent effort, the Greater Arab Free Trade Area, will be a critical step in determining the future of economic integration in Arab countries.

[5] See the Cooperation Council Charter, List of Member States (1981) available at <http://www.gccsg.org/eng/index.php?action=Sec-Show&ID=1> (last visited Sep. 13, 2010).

[6] See Stacy Amity Feld, Language and the Globalization of the Economic Market: The Regulation of Language as a Barrier to Free Trade, 31 Vand. J. Transnat'l L. 153, 158 (1998) (discussing how language must be recognized and reckoned with as an important factor of the economy. Economic integration cannot be viewed independent of cultural or sociopolitical factors).

It might be assumed that the U.S. participated actively in the birth of the GCC. Part of the reason for U.S. involvement was the desire to stem Islamic extremism and Iran influence in the Gulf region.[7] Also, the U.S. had a vision of an international economic system conducive to free trade and unfettered investment. In modern times, the U.S. President announced his initiative to create a Middle East Free Trade Area by 2013.[8] This initiative is designed to deepen U.S. trade relationships with all countries of the region, through steps tailored to individual countries' level of development. Since that announcement, the U.S. has concluded FTA negotiations with Bahrain and has signed Trade and Investment Framework Agreements (TIFAs) with Saudi Arabia, the United Arab Emirates, Kuwait, Qatar, Yemen, and now with Oman.[9] With the U.S. as a trade partner, its involvement exhibits a desire for peace, stability, and economic opportunity in the Gulf region.

The idea of regional integration gained ground by signing the GCC Charter. It envisioned a closer economic relationship between member states. The aim of the GCC was to promote cooperation in all fields of economic activity in order to increase and maintain economic stability, fostering closer relations among its members, and contributing to the progress and development of the Gulf region.[10] The other founding documents that established the GCC, its main organizations, and its executive procedures are the Supreme Council Rules of Procedure, the Ministerial Council Rules of Procedure, and the Commission for the Settlement of Disputes Rules of Procedure.

---

[7] See Gawdat Bahgat, American Policy in the Persian Gulf, 24 Fletcher F. World Aff. 143, 153-155 (2000) (for the last two decades, Iran has presented the United States government with a difficult dilemma, leading to two conflicting pillars in American foreign policy: a desire to promote Washington's economic interests around the world and a determination to exert pressure on Tehran to change its behavior on certain foreign policy issues with respect to the Middle East peace process, the sponsoring of international terrorism and attempts to acquire and develop weapons of mass destruction).

[8] See Mike Allen & Karen DeYoung, Bush Calls Trade Key to Mideast; President Launches Plan for U.S. Pact in Region, Wash. Post A01 (May 10, 2003). See also Grary G. Yerkey, President Bush Lays Out Broad Plan for Regional FTA with Middle East by 2013, 20 Intl. Trade Rep. (BNA) 856 (May 15, 2003) (the U.S. will employ a "building-block" approach. This approach requires, as a first step, a Middle East country to accede to the WTO or concluding Trade and Investment Framework Agreement(s) (TIFA). Afterward, the U.S. will negotiate FTA with individual countries. Finally, preferably before 2013, a critical mass of bilateral FTAs would come together to form the broader US-Middle East FTA).

[9] See Rossella Brevetti, House Approves by Large Margin U.S.-Bahrain Free Trade Agreement, 22 Int'l Trade Rep. (BNA) 2031 (December 15, 2005). See Gary G. Yerkey, U.S. Free Trade Talks with UAE on Track to be Completed by Year-end, 23 Intl. Trade Rep. (BNA) 326 (March 2, 2006).

The European Union (EU) is often touted as the example to follow as far as the GCC is concerned. However, one must remember that both GCC and the EU were created with completely different objectives and functions.[11] The differences between the GCC and EU are great that comparison in this regard between them is not productive. Thus, the article will examine solely the GCC as an economic integration block. The article is divided into two major sections. This article first discusses the governance structure of the GCC. The article, then, focuses on the progress and the issues surrounding the GCC integration initiatives.

**II. Governing and Institutional Architecture**

Given the logistical problems that could turn into significant political problems of a two-tiered system- member states and economic integration institutions- the quality of the GCC governance structure is most important. Quality can be defined as enough structure to define the institutions and to allow those institutions serious, actual authority to serve as a counterweight to the national interests likely to otherwise stifle governance.

---

[10] The aim of the establishment of the GCC can be deduced from the Charter's preamble: "to effect co-ordination, integration, and interconnections between them in all fields". See the Cooperation Council Charter, Preamble (1981) available at <http://www.gccsg.org/eng/index.php?action=Sec-Show&ID=1> (last visited July. 20, 2010).

[11] One main difference in the formation of the GCC and EU is that the founding members of the EU united to build each other back up from the ravages of World War II. Their economies were wrecked, and in turn, each contributing nation found strength in partnership with each other. In contrast, GCC member states were in very different positions when the GCC was formed. None of these countries were seeking to rebuild their economies or, more importantly, seeking political integration. Instead, GCC member states united to seek the advantages a trade block would bring each other. The advantages were purely economic. The founding members of the EU were interested in much more than economic integration. The European nations were also concerned with surrendering some of their independent sovereignties to further political integration. This political integration, along with economic integration, could also be viewed as providing strength and protection in case any threat of war was to occur on European lands in the future. Member states of GCC expressed clear unwillingness to surrender their individual sovereignties such as the lack of an accession provision for new members. See Desmond Dinan, Fifty Years of European Integration: A Remarkable Achievement, 31 Fordham Int'l L.J. 1118, 1123-1129 (2008). See also John A. Sandwick, The Gulf Co-operation Council: Moderation and Stability an Interdependent World 71-72 (Westview Press, 1987).

The GCC governing structure is composed of the Supreme Council, the Ministerial Council, the Secretariat-General, and the Commission for Settlement of Disputes.[12] These entities have the authority to establish any sub-agencies when necessary.[13]

**A.     The Supreme Council**

The Supreme Council is the most powerful GCC institution and is the head of the GCC governance structure. The Supreme Council is composed of the head of each of the member states.[14] Its presidency rotates among the member states in alphabetical order. The Supreme Council meets annually and the validity of any of its meetings is dependent upon the attendance of two-thirds of the member states.[15]

The Supreme Council is the principal legislative body of the GCC and authorizes the other GCC entities to implement its decisions in pursuit of its mandate to realize the objectives of the GCC. For example, the Supreme Council has the power to amend the Charter.[16] In addition, the Supreme Council reviews the matters of interest to the member states, establishes the higher policy for the GCC, reviews the recommendations and reports submitted by the Ministerial Council for approval, reviews the reports prepared by the Secretary-General, approves the bases for dealing with other states and international organizations, nominates the members of the Commission for the Settlement

---

[12] See the Cooperation Council Charter, *supra* note 10, at art. 6.
[13] *Id*.
[14] *Id*. art. 7.
[15] *Id*.
[16] *Id*. art. 8.

Disputes, and appoints the Secretary-General.[17] The authority of the Supreme Council is a general one but not is very detailed such as in the case of conflict management and other emergencies.

The voting of the Supreme Council decides whether or not a particular resolution is to be adopted and thus binding on the member states. In the Supreme Council, each member has one vote.[18] The Charter divides the voting into two kinds, substantive matters and procedural matters.[19] On the one hand, substantive matters must be approved consensus.[20] In other words, a particular decision is adopted if each and every member state does not veto that decision. Unfortunately, consensus is the first step toward disempowering decision-making bodies.[21] On the other hand, procedural matters must be approved by a majority vote.[22] The rationale for dividing matters into substantive and procedural indicates that the stakes are high with substantive matters and as such subject to consensus decision-making.

There is no clear reference in any of the GCC agreements as to what should be considered a substantive matter or a procedural matter. In the absence of such a clear reference, the Supreme Council can define the meaning of substantive and procedural matters. The Rules of Procedure of the Ministerial Council granted the Council the authority to decide what is substantive and what is procedural.[23] Although, the Rules of Procedures of the Ministerial Council applies to the Ministerial Council, it can be argued that by analogy the same could apply to the Supreme Council.

---

[17] *Id*.
[18] *Id*. art. 9.
[19] This is somewhat is similar to the European Union. Under the Luxembourg Accord, the member-states agreed to regard matters that might otherwise be approved by majority vote as subject to a consensus. See P.S.R.F. Mathijsen, A Guide to European Community Law (5th ed. 1990). See also Wolf Sauter, The Constitution of the European Union, 4 Colum. J. Eur. L. 27, 33 (1998).
[20] See the Cooperation Council Charter, *supra* note 10, at art.9.
[21] The Luxembourg Accord of the EU stands as a paradigm example of paralysis in decision-making when consensus is required. Other examples include the GATT process; though majority decision-making was the rule, few decisions were taken by that rule and instead consensus became the de facto reality. The ineffectiveness of the GATT's moribund decision-making, which was a consensus process, led to the changes in that area. See John Jackson et al., Legal Problems of International Economic Relations 215 (2002).
[22] See the Cooperation Council Charter, *supra* note 10, at art.9.

**B. The Ministerial Council**

The Ministerial Council is composed of the foreign ministers of the member states or other designated Ministers.[24] In other words, delegates from the appropriate ministries may attend depending upon the issue at hand. The Ministerial Council convenes regular meetings every three months,[25] but it can also convene extraordinary meetings based on an invitation by one of the member states. Like the Supreme Council, the Ministerial Council's meetings are considered valid if attended by two-thirds of the member states.[26]

The powers of the Ministerial Council are more detailed than the Supreme Council. These powers include proposing policies, prepare recommendations, studies and projects aimed at developing cooperation and coordination between member states in various fields; endeavoring to encourage, develop, and coordinate activities existing between member states in all fields; developing existing cooperation between member states' industry and Chambers of Commerce, and encouraging the movement of workers who are citizens of the member states within the GCC; and reviewing matters referred to it by the Supreme Council.[27] In summation, in terms of power, the Ministerial Council can actually make its own decisions but in other cases can be subject to superior approval.

---

[23] The Rules of Procedures of the Ministerial Council states that if there is any disagreement among the member states to decide what is substantive and what is procedural, the matter shall be settled by a majority vote. See Rules of Procedure of the Ministerial Council, art. 33.2 (May 25, 1981), available at <http://www.gcc-sg.org>. This means that the matter to settle what is substantive and what is procedural is considered a procedural matter in itself.
[24] See the Cooperation Council Charter, *supra* note 10, at art.11.
[25] *Id*.
[26] *Id*.
[27] *Id*. art. 12.

Voting in the Ministerial Council is addressed in both the Charter and the Ministerial Council Rules of Procedure. Like the Supreme Council, each member in the Ministerial Council has one vote.[28] Again both agreements divide voting into two categories, substantive matters and procedural matters. Like the voting rules for the Supreme Council, resolutions on substantive matters must be approved unanimously and resolutions on procedural matters must be approved by the majority.[29] However, unlike the voting rules for the Supreme Council, the Ministerial Council Rules of Procedure explicitly has the authority to resolve the issue of substantive and procedural.

**C. The Secretariat General**

The Secretariat is composed of a Secretary General and other necessary assistants and staff members.[30] The Secretary General should be a citizen of the GCC and is appointed by the Supreme Council for a period of three years, which may only be renewed once.[31] The number of professional civil servants who work in the GCC Secretariat remains unclear.[32] However, one anticipates that the Secretariat consists of primarily trade economists and trade lawyers.

The capacity of the Secretariat to affect outcomes, however, depends on what it actually does and the saliency of the issues at stake. The Secretariat performs primarily three functions in the GCC process: the organization of meetings and recording of minutes; research on GCC issues; and follow up the implementation by member states of

---

[28] *Id*. art. 13.
[29] *Id*. art. 13. See also Rules of Procedures of the Ministerial Council, *supra* note 23, art. 33.1.
[30] See the Cooperation Council Charter, *supra* note 10, at art. 14.
[31] *Id*.
[32] The total size of the WTO Secretariat is around 630 persons, a large proportion of whom are translators and support staff, compared to a World Bank staff of around 10,000 and an International Monetary Fund staff of around 2,700. See WTO, Annual Report 2005, at 105 (2005); The World Bank Staff, <http://worldbank.org> (follow "About" hyperlink; then follow "Staff" hyperlink) (last visited Feb. 3, 2009); The IMF at a Glance, <http://www.imf.org> (following "For Journalists" hyperlink, then follow "IMF at a glance" hyperlink under Facts and Issues) (last visited April 14, 2010).

Supreme Council and Ministerial Council resolutions.[33] Although not explicitly stated, the secretariat can mediate between member states and liaise with international organizations.[34] On the basis of their mandate, the Secretariat members, at least at the margins, help shape knowledge, frame issues, identify interests, facilitate coalition-building, and thereby affect outcomes. Thus, the GCC Secretariat appears to have no substantive decision-making power and works as an intermediary to facilitate state-to-state discussions, negotiations and monitoring. The Secretariat role can be viewed as one of servicing negotiations and oversight of obligations. The member states of GCC have clearly sought to ensure that control over the integration process remains in the hands of the member states and not in a group of independent civil servants.

The Secretariat must maintain a reputation for impartiality at all times.[35] The Secretariat is usually cautious not to appear partisan with no direct and clear preference. Thus, for example, the Secretariat is discouraged from offering counsel as to how an obligation could be interpreted by a member state to facilitate its objective(s).

The Secretariat could influence outcomes through its research for instance. Member states expect the Secretariat to keep abreast of trade and economic studies, particularly those conducted by other international organizations. In distributing information to all interested parties, the Secretariat helps create a common base of understanding. Upon request, the Secretariat researches and prepares papers on specific issues. By 2010, the Secretariat had provided members with fifty-eight papers, totaling hundred of pages.[36]

---

[33] See the Cooperation Council Charter, *supra* note 10, at art. 15.
[34] This can be deduced by the language of article 15 which entrusts the Secretariat with "any" other task assigned to it by the Supreme Council or the Ministerial Council. *Id*.
[35] *Id*. art. 16.
[36] See GCC Library English Publication, available at <http://library.gcc-sg.org/English> (last visited Sept. 10, 2010).

Secretariat submissions addressed, among other matter, trade liberalization, economic implications of trade liberalization, currency, environment, and regional cooperation. The Secretariat's reports are largely informative and not argumentative in tone.

**D. The Commission for Settlement of Disputes**

Perhaps the most important aspect of the GCC's governance structure is the dispute settlement body and process. The dispute settlement body and process play a crucial role especially at times when the political will of the integration is questionable. If the dispute settlement body is seen independent and able to ascertain its power, this will engender confidence in an integration scheme. In the North American Free Trade Area (NAFTA) integration example, panels played an important role in strengthening the free trade area.[37] The binding rulings of NAFTA panels helped clarify valid regulations and policies of member states, and they set aside those that did not conform to the obligation to liberalize trade.[38]

The GCC Charter establishes the Commission for Settlement of Disputes (Commission). The Commission is composed of at least three citizens of the member states.[39] Panelists of the Commission are not appointed full-time or permanent rather they are selected on an *ad hoc* basis.[40] As presently drafted, the Charter does not provide any guidelines for the selection of panelists of the Commission in terms of their qualifications, age, or years of expertise in the area of trade law, policy, or economics either in the domestic or international arena. Additionally, the Charter does not indicate whether panelists of the Commission would be expected to act in their governmental capacity or neutrally. Panelists should be required to serve in their individual capacities and not as government representatives, nor as representatives of any organization to avoid any political or other undue influence.

---

[37] Chapter 19 of NAFTA establishes binational panels to review final determinations of antidumping and countervailing duty measures imposed under NAFTA parties' national antidumping and countervailing duty law. Binational panels are also permitted to review amendments to the antidumping and countervailing duty laws of the NAFTA parties. For more discussion of NAFTA's dispute resolution processes under chapter 19 procedure see David S. Huntington, Settling Disputes Under the North American Free Trade Agreement, 43 Harvard International Law Journal 407 (1993). See also Myung Hoon Choo, Dispute Settlement Mechanisms of Regional Economic Arrangements and Their Effects on the World Trade Organization, 13 Temp. Int'l & Comp. L.J. 253, 254 (1999) (arguing that the achievement of "diverse policy objectives, i.e., trade liberalization or environmental protection, as well as true harmonization among nations" is contingent upon the dispute settlement mechanisms available under the particular regional free trade agreement).

[38] See NAFTA Panel Ruling, Tariffs Applied by Canada to Certain U.S.-Origin Agricultural Products, Dec. 2, 1996, CDA-95-2008-01, NAFTA Panel Ruling, U.S. Safeguard Action Taken on Broomcorn Brooms from Mexico, Jan. 30, 1998, USA-97-2008-01.

[39] The Commission has its seat in Riyadh, the capital of Saudi Arabia. See Rules of Procedure of the Commission for Settlement of Disputes, art. 4(a) (May 25, 1981), available at <http://www.gcc-sg.org>.

The Commission has jurisdiction to consider matters referred to it by the Supreme Council regarding disputes between member states as well as disputes over the interpretation and implementation of the Charter.[41] Therefore, the Member State may bring an action before the Commission alleging the failure of another member State to fulfill the obligations of the Charter, and in these matters the Commission has original jurisdiction. The jurisdiction of the Commission may also extend to review decisions or actions of the Supreme Council or the Ministerial Council for consistency with the Charter. However, the ability of the Commission to hear disputes between Member States depends upon discretionary and consensus "referral" by the Supreme Council. Moreover, neither the Charter nor the Rules of Procedure of the Commission have any provision for individual access, direct or otherwise, to the Commission. Moreover, there is no provision regarding the interaction between the Commission and the various national courts systems and role of the latter in enhancing regional economic integration. Another problematic issue is that the Charter is silent about situations where the actions or domestic law of a particular Member State - while not in violation of the GCC agreement - may nevertheless inadvertently contradict or nullify the purposes of the agreement. At any rate, the ultimate authority of the Commission is a subject for future observation and certainly a strong argument can be made for adopting these concepts at some point in the future.

---

[40] *Id*. See also the Cooperation Council Charter, *supra* note 10, at art. 10.
[41] If a dispute arises over interpretation or implementation of the Charter and such dispute is not resolved within the Ministerial Council or the Supreme Council, the Supreme Council may refer such dispute to the Commission for the Settlement of Disputes. See Rules of Procedure of the Commission for Settlement of Disputes, *supra* note 39, at art. 3. See also the Cooperation Council Charter, *supra* note 10, at art. 10.

Regarding the recommendations and opinions issued by the Commission, the Rules of Procedure of the Commission sets out four guiding points when issuing such recommendations. First, the recommendations or the opinions should be in accordance with the Charter, international laws and practices, and the principles of Islamic Shariah.[42] Second, the Commission, while considering any dispute and before issuing the final recommendation, may ask the Supreme Council to take an interim action called for by circumstances.[43] Third, the Commission should justify its recommendations by specifying the reasons on which they were based.[44] Finally, if the opinion is not issued unanimously, the dissenting members are entitled to record their dissenting opinion.[45] The Commission's task is considered to be completed upon the submission of its recommendations or opinions to the Supreme Council.[46]

The Charter and Rules of Procedure of the Commission do not include any language for appellate review process.[47] The Charter and Rules of Procedure do not discuss whatsoever the scope of the "review" in terms of factual matters, legal substance, or the rules and nature of its proceedings. Furthermore, the scope of the review is not determined. Thus, because the GCC agreement's dispute settlement mechanism lacks depth and specificity, modifications that would significantly enhance the mechanism's value, and infuse greater certainty into the dispute settlement process are required.

---

[42] *Id*. art. 9.
[43] *Id*.
[44] *Id*.
[45] *Id*.
[46] *Id*. art. 4.
[47] See Nobuo Kiriyama, Institutional Evolution in Economic Integration: A Contribution to Comparative Institutional Analysis for International Economic Organization, 19 U. Pa. J. Int'l Econ. L. 53, 67-68 (1998) (emphasizing the role of effective appellate review processes in ensuring predictability and the confidence of the parties).

To date, the GCC dispute settlement process has not been used. Lack of use of the dispute settlement process is not likely to form an intelligible body of jurisprudence providing parties with reliable and consistent interpretations of rules. Further, there are several lacunae not addressed by the dispute settlement mechanism. There are no elaborate and detailed criteria for the composition of the panels. Another problem is the ambiguity in the scope and jurisdiction of the panel, which could be a major threshold issue in determining when and what disputes member countries could refer for resolution. Another obstacle is the lack of procedures for the operation of the panel, as well as the largely undefined qualifications of its members. In addition, the appellate review process is absent as there is lack of scope and the procedure for review of the legal versus substantive matters. Still more issues are left open-ended such as the enforcement of decisions, procedures for withdrawing and reinstating concessions.

Model Rules of Procedures should be developed so as to determine the numbers of panelists, their qualifications, expertise, nationality, and remuneration. Model rules of procedures may include policies, practices, and procedures for receiving initial and rebuttal written submissions, how oral hearings will be conducted before a panel, and mandatory time limits for each step. The dispute settlement should call for increased transparency in proceedings, in particular the opening up of panel hearings to the public.

In regards to the presentation of confidential business information in the panel proceedings, portions of any dispute hearing dealing with such confidential information would not be open to the public.

The Model Rules of Procedures should have an elaborate system of sanctions and measures in order to enforce trade norms. The most salient feature of dispute settlement under the GCC should be the possibility of authorizing a trade sanction such as the suspension of tariff reductions against a scofflaw member for non-compliance. Trade sanctions or threat thereof are to taken to ensure that the Arab country in breach brings its practices into conformity. There can be other alternatives for trade sanctions. For example, instead of trade sanctions, any GCC Member State guilty of illegal trade practices could pay a fine equal to the value of the damages assessed. Other alternative can be membership sanctions that limit or deny privileges of membership for any GCC Member State that fails to comply with the provisions of the agreements. Among the membership benefits that can be withdrawn are the right to vote and the ability to obtain financial or technical assistance. The goal of these sanctions and measures is to fortify the agreement rules and promote respect for them.

**III. Expanding the GCC into a Customs Union, Common Market, and Monetary Union**

The integration of the GCC has followed a progression from one stage to another while developing its own sense of political structure, its own legal system in relation to its member states, and its own pro-business trading environment. Of course, this integration is not one that could be formulated and implemented within even a few years. Rather, it must be viewed as a long-term process in which there are gradual transitions and many intermediate steps that are needed along the way. The member states of the GCC can be seen as fully integrated economically and even more so with the projected adoption of their common currency, yet to be named.[48] In addition, the GCC continues to change and evolve by enacting new agreements that further strengthen the member states' ties to each other. There are other integration efforts beyond those technically falling as customs union, common market, and monetary union. Agreements of cooperation and sector consultation are part of the Gulf integration process.[49] As the following sections point out that further integration by the members of the GCC will continue to cause the members to surrender some of their national sovereignty to a "supranational authority."

**A. The GCC Customs Union**

The GCC Free Trade Area continued for almost twenty years until the end of 2002 when it was replaced by the GCC Customs Union. On January 1, 2003, the GCC Customs Union took effect marking a quality shift in the joint economic actions between its member states. A customs union creates a wider trading area, removes obstacles to competition, makes possible a more economic allocation of resources, and thus operates to increase production, and raise planes of living.

The most distinct difference between a customs union and a free trade area (FTA) is that in an FTA, countries are free to set their own external trade policy; whereas in a customs union, members as a whole set a common external policy. The biggest advantage of a customs union is that, because members have a common external tariff, it facilitates deeper integration and allows the members to have simpler internal border formalities, possibly none at all.[50] In contrast, an FTA leaves external trade policy to individual member governments and faces a problem known as trade deflection or transshipment.[51] That is the problem of the redirection of imports from outside countries through the FTA member with the lowest external tariff in order to exploit the tariff differential. The usual solution is rules of origin--the apparently reasonable requirement that goods qualifying for tariff-free trade should be produced in a member country rather than just passing through that country. Countries in an FTA retain the sovereign power to decide individually whether, and to what level trade restrictions should be imposed on nonmembers. The level of economic and political integration required to establish an FTA is not as extensive as a customs union. Thus, an FTA attracts those states preferring a loose-knit regional structure.[52] A customs union agreement, however, does not necessarily imply an overt surrender of national sovereignty. On the other hand, establishing identical tariff barriers against imports from nonmembers requires a commitment to common decision-making, weakening the ability of participating countries to determine national trade policies independently.

---

[48] See Disunited Arab Emirates; Monetary union in the Gulf, The Economist (May 23, 2009).
[49] See Economic Agreement, *infraa* note 59, at Chapters IV, V, VI, and VII.
[50] See The World Bank, Trade Blocs 75 (2000).

Article XXIV of GATT embodies the perception that genuine customs unions and free trade areas are congruent with the Most-Favored-Nation principle and that regional trade agreements ought to be building blocks toward multilateralism.[53] A customs union eliminates barriers to trade in goods between or among its member and adopts a common external tariff that all members of the customs union apply to trade from countries outside the union. A customs union is defined in GATT Article XXIV: 8 as the substitution of a single customs territory for two or more customs territories, so that duties and other restrictive regulations of commerce are eliminated with respect to "substantially all the trade" between the constituent territories of the union or at least with respect to substantially all the trade in products originating in such territories,[54] and substantially the same duties and other regulations of commerce are applied by each of the members of the union to the trade of territories not included in the union.[55] In addition, Article XXIV: 5 mandates additional requirements by stating that duties and other regulations of commerce imposed at the institution of any such union or interim agreement in respect of trade with contracting parties not parties to such union or agreement is "not on the whole be higher or more restrictive than the general incidence" of the duties and regulations of commerce applicable in the constituent territories prior to the formation of such union.[56] The language used for these requirements in Article XXIV of GATT, however, has long been criticized for its ambiguity and inconclusiveness.

---

[51] See Zakir Hafez, Weak Discipline: GATT Article XXIV and the Emerging WTO Jurisprudence on RTAs, 79 N. Dak. L. Rev. 879 (2003).
[52] See Richard Gibb, Regionalism in the World Economy, in Continental Trading Blocs: The Growth of Regionalism in the World Economy 25 (Richard Gibb & Eieslaw Michalak eds., 1994).
[53] See Report of the Panel, Turkey-Restrictions on Imports of Textile and Clothing Products, WTO Doc. WT/DS34/R, para. 2.2 (May 31, 1999).

The establishment of a customs union requires a common external tariff scheme, but permits constituent members of customs unions flexibility in implementing that scheme.[57] Members of a customs union need only apply "substantially the same" duties and commerce regulations to countries outside the customs union. Additionally, execution of this common tariff scheme must not raise barriers to trade with countries outside of the customs union.

---

[54] Exactly what percentage of trade constitutes a "substantial" amount has never been defined in bright-line terms. Discussions in GATT Working Parties have centered on whether the concept of "substantial' should be understood in qualitative terms (no exclusion of major sectors) or in quantitative terms (percentage of trade of the members covered). See World Trade Organization, Analytical index: Guide to GATT law and Practice, Vol. 2, 824-27 (1995). The Appellate Body in the Turkey's Restriction on Textile affirmed the ordinary meaning of the term "substantially" in the context of sub-paragraph 8(a) appears to provide for both qualitative and quantitative components. See Appellate Body Report on Turkey's Restriction on Textile, Appellate Body Report, Turkey-Restrictions on Imports of Textile and Clothing Products, WTO Doc. WT/DS34/AB/R, paras. 49-50 (Oct. 22, 1999).
[55] See Annexes to GATT Article XXIV: 8(a),, reprinted in Raj Bhala, International Trade Law Handbook 250-54 (2d ed. 2001).
[56] The evaluation under paragraph 5(a) of Article XXIV of the general incidence of the duties and other regulations of commerce applicable before and after the formation of a customs union shall in respect of duties and charges be based upon an overall assessment of weighted average tariff rates and of customs duties collected. This assessment shall be based on import statistics for a previous representative period to be supplied by the customs union, on a tariff-line basis and in values and quantities, broken down by WTO country of origin.

The basis of customs law in the GCC is the Common Customs Code of the GCC.[58] The Code provides a uniform set of general rules to be implemented by national customs authorities to harmonize the application of duties and procedures for processing imports into the GCC. The Code mandates the implementation of a common external tariff or common customs tariff (CCT) scheme applicable to all third country imports by the Member States, as well as requires cooperation between national customs authorities in relation to customs matters.[59] The common external tariff of the GCC Customs Union is 5 percent.[60] The common external tariff is a flat-rate charge thus erecting a single tariff wall which no individual state is free to breach.

The CCT is comprised of the tariff classification system embodied in the International Convention on the Harmonized Commodity Description and Coding System (Harmonized System) and the Common Customs Tariff.[61] The Harmonized System is a multipurpose international product nomenclature developed by the World Customs

---

[57] Members need not apply "the same duties and other regulations of commerce as other constituent members with respect to trade with third countries. See Annexes to GATT Article XXIV, Bhala, *supra* note 52.

[58] See the Common Customs Code of the GCC States (January 2003), available at < http://library.gcc-sg.org/English/Books/customs2003.htm > (last visited June 19, 2010).

[59] *Id*. art. 9. The Economic Agreement of the GCC countries state that member states shall establish uniform minimum Customs tariffs applicable to the products of countries other than GCC member states. See Economic Agreement, art. 4.1, (Dec. 31, 2001), available at <http://library.gcc-sg.org/English/econagreeeng2003.htm>. The Economic Agreement amended and revised the Unified Economic Agreement, which was signed and approved on November 11, 1981. See Unified Economic Agreement (Nov. 11, 1981), available at <http://www.gcc-sg.org/Economic.html>.

[60] See Implementation Procedures for the Customs Union of the GCC (2003), available at <http://www.gccsg.org/eng/index.php?action=Sec-Show&ID=93 > (last visited June 19, 2010).

[61] See International Convention on the Harmonized Commodity Description and Coding System, June 14, 1983, 1035 U.N.T.S. 3. (Entered into force Jan. 1, 1988).

Organization (WCO). The Harmonized System comprises about 5,000 commodity groups, each identified by a 6-digit code, arranged in a legal and logical structure and is supported by well-defined rules to achieve uniform classification. The Harmonized System is used by more than 190 countries and economies as a basis for their customs tariffs and for the collection of international trade statistics.[62] The CCT is determined in Common Customs Code of the GCC, and consists of several distinct elements.[63] These include the Combined Nomenclature, preferential tariff provisions between the GCC and third countries, and measures providing for a reduction in, or relief from, import duties.

**B. The GCC Common Market Policy**

The GCC common market was created in 2008 by the decision made by the GCC leaders at their summit, which was held on December 3, 2007, and was meant to create a common market among its six members.[64] This common market would include the graduated elimination of all customs duties among its members, the creation of a common external tariff, the adoption of a common trade policy, and the harmonization of economic policies. Moreover, the common market would have to adopt many directives and resolutions seeking to eliminate barriers to free trade and to harmonize the legal and regulatory systems of the member states, as well as to form the basis of a system of community law.

The GCC common market would comprise trade that would promote the free movement of capital, goods, services, and labor. The Economic Agreement of the GCC obliges each Member State to accord other Member States' natural and legal citizens the same treatment,[65] which includes: free movement of goods,[66] free movement of citizens, right of residency,[67] free movement of capital,[68] the right to work in private and government jobs,[69] the right to be engaged in all professions and crafts,[70] the right to be engaged in all economic, investment, and service activities,[71] and the right to own stock and form corporations.[72] Each of these freedoms and rights are crucial if the goal of a regional integration is to break down all distinctions among states in the area of economic activity. As a result, the common market would involve a deeper level of integration within the GCC member states.

---

[62] See Position of Contracting Parties to the Harmonized System Convention as of March 31, 2010, available at
<http://www.wcoomd.org/files/1.%20Public%20files/PDFandDocuments/HarmonizedSystem/Situation_as of31Jan2010.pdf> (last visited August 5, 2010).
[63] See the Common Customs Code of the GCC States, *supra* note 58, at arts. 19, 67, 69, 77, 89, 95, and 98.
[64] See Gulf Countries Launch Common Market, Jordan Times (Dec. 5, 2007).

Does a common market require a common law? Fundamental to the formation of an integrated GCC is the creation of a common legal system. A common market cannot operate smoothly without certain generally recognized rules and procedures, without a core of common legal institutions and convictions.[73] Harmonization of laws facilitates transactions concluded between different sovereignties and encourages all forms of economic exchange.[74] Having taken this into consideration, and in order to achieve its objectives and aims of establishing a regional block, the GCC attempted to harmonize various laws and regulations. For example, the GCC adopted Trade Marks Regulations,

---

[65] See Economic Agreement, *supra* note 59, at art. 3.
[66] Synergies are created by movement of goods. However, one has to acknowledge the reality of GCC economy based on the export of primary products (oil) and thus unstable terms of trade and limited economies of scale for production, especially industrial production. The efforts of economic integration should address these factors through economic policies that flow from integration.
[67] *Id*. art. 3(1).
[68] *Id*. art. 3(7).
[69] *Id*. art. 3(2).
[70] *Id.* art. 3(4).
[71] *Id*. art. 3(5).
[72] *Id*. art. 3(9).
[73] See Frederick Abbott, Regional Integration and the Environment: The Evolution of Legal Regimes, 68 Chi.-Kent L. Rev. 173, 189 (1992).
[74] Harmonization is seen as a transitive verb. There is a harmonizer - the agent of harmonization- and there is an object of harmonization - the diverse national laws which must undergo a process of change or transformation to reduce or eliminate their differences. Martin Boodman, The Myth of Harmonization of Laws, 39 Am. J. Comp. L. 699, 707 (1991).

Commercial Agencies Regulations, Common Regulations of Patent, Model law for Foreign Investment, Common Commercial Law, Law of Commercial Registry, Common Marine Law, Law of Supervision on Insurance, Unified Law of Companies, and the Unified Law of Personal Status.[75] These common laws and many others are non-obligatory laws. In other words, member states are not obliged to implement them.

Conflict between different legal regimes can be mitigated through a process of unification of laws, through adoption of uniform rules that are devised by and adopted by separate states. Harmonization of legal rules in the GCC member states has occurred through voluntary change - not imposed by a supranational organization - but spontaneously by independent national institutions, in order to respond to the GCC environment. In order to be in harmony, changes are made, but they are voluntary rather than imposed. For example, model laws are devised for submission to the legislatures within the GCC member states.

The GCC common market would help make the Gulf region more competitive globally.[76] The perception was that GCC's common market major advantages would be economic through increased trade with fewer barriers and economic complementarity, and that the impact of these changes would be positive, both short- term and long-term.[77] It would result in a larger, more protected market for the area's products and GCC countries would be better off economically.

---

[75] This is not an exhaustive list. More details are available in Cooperation Council for the Arab States of the Gulf Secretariat, Achievements Synopsis of the Arab States of the Gulf (2009) (in Arabic), available at <http://library.gcc-sg.org/book8a01e.htm> (last visited Sept. 1, 2010)
[76] See Gulf Countries Launch Common Market, *supra* note 64.
[77] *Id*.

Between 1983 and 2008, trade among the GCC member states increased dramatically.[78] Trade and investment between GCC member states had quadrupled, and a substantial interdependence between the economies of these countries had developed.[79] Indeed, at this point, things were going so well that the idea of a monetary union was being discussed.

**C. The Monetary Union**

Monetary union has always been considered to be the next step to economic union in the process of economic integration.[80] This concept was already in the original text of the Economic Agreement which discusses currency unification and monetary union according to a specified timetable.[81] This unification is to be achieved through the harmonization of economic and monetary policies, including banking legislation.[82] Thus, there is an obligation to arrive at a common monetary policy and creating a common currency.

The monetary union process began in December 2001.[83] The monetary union has two main components: 1) the establishment of the Monetary Council and GCC Central Bank for the purpose of exercising the primary monetary powers; and 2) the creation of a single currency to replace national currencies as the sole legal tender for the member states in the GCC.[84] It is important to ensure that the GCC Central Bank will have total independence in its decision-making. This point represents a crucial policy decision because GCC central banks were not independent of their respective governments at the time the Monetary Union Agreement was approved.[85] If the GCC Central Bank is granted full independence it can formulate a single monetary policy no matter how unpopular and without regard to political pressure from the participating member states.[86] This full independence is an important feature of the GCC monetary union to which member states have to surrender their sovereignty in order to reap the anticipated benefits of the monetary union.

---

[78] The volume of Intra-GCC trade increased from less than US $6 billion in 1983 to some US $20 billion in 2008 i.e. an increase by 75.5% over the past ten years or an average annual growth accounting for some 7.5%. After the formation of the common market in January 2008, the volume of Intra-GCC trade has increased by an annual growth average that exceeded 20%. See Cooperation Council for the Arab States of the Gulf Secretariat, *supra* note 75, at 241.
[79] *Id*.
[80] Regional integration tends to involve six stages of development: (1) preferential trade arrangement; (2) free trade area; (3) customs union; (4) single market; (5) monetary union; and (6) political union. See Bela Balassa, The Theory of Economic Integration 2 (1962). See also Myung Hoon Choo, Dispute Settlement Mechanisms of Regional Economic Arrangements and their Effects on the World Trade Organization, 13 Temp. Int'l & Comp. L.J. 253, 255 (1999).
[81] See Economic Agreement, *supra* note 59, at art. 4. The single currency is to be created by 2010, as voted on by the Cooperation Council's Member State in their Summit on December 30 - 31, 2001. See B. Laabas and I. Limam, Are GCC Countries Ready for Currency Union, Arab Planning Institute, Kuwait, at 2, available at < http://www.arab-api.org/wps0203.pdf> (Apr. 2002).
[82] See Economic Agreement, *supra* note 59, art. 4.
[83] Unlike the customs union whereby the member states agreed to a deadline for the implementation of the customs union and stated this deadline in the Economic Agreement, the monetary union does not include such a deadline. Although, the deadline for the establishment of the monetary union was not stated in the Economic Agreement, the member states voted on a deadline of 2010.

To achieve a single currency, there must be an interim period that serves as a period designed to ensure a smooth changeover from national currencies to the single currency. During this transitional period, businesses and commercial enterprises throughout the GCC can be "encouraged "to use the single currency for their accounting and transactions. At first, the GCC Central Bank must adopt a policy of "no prohibition, no compulsion" use of the single currency. For instance, internal and cross-border payments made by crediting an account can be made either in the single currency or national currency units. In addition, many businesses could adopt voluntary rules of conduct for incorporating the single currency during the transitional period.

---

[84] See Cooperation Council for the Arab States of the Gulf Secretariat, *supra* note 75, at 92.
[85] See GCC Central Banks, available at <http://www.centralbank.ae/gcc_central_banks.php> (last visited May 5, 2010).
[86] One of the functions of the Central Bank is to conduct the monetary policy. The Central Bank primarily does this through regulation of the banking industry. The Central Bank controls the supply of reserves available to banks through the purchase and sale of Treasury bills and through its administration of the discount window and discount rate. The Central Bank has the power to create money by purchasing Treasury bills on the open market. The Central Bank can contract the money supply by selling Treasury bills on the open market. Creating too much money will cause inflation; contracting the money supply too much will lead to recession. The same logic applies to the discount rate. When the discount rate is high, banks are discouraged from borrowing, thus somewhat retracting the economy. However, when discount rates are low, banks are encouraged to borrow money and lend it to the public, thus increasing the money supply. See John W. Head, Getting Down to Basics: Strengthening Financial Systems in Developing Countries, 18 Transnat'l Law. 257, 260-266 (2005).

The introduction of the single currency will not only affect the governmental and business sectors but also the general population.[87] The general population should become familiarized with the single currency during the transitional period. For instance, merchants could price their goods in the single currency and national currency in order to familiarize consumers with the value of the single currency. Moreover, banks could serve their clients' deposit and loan accounts on a dual currency basis. It is difficult to gauge the percentage of the population that will use the single currency. Therefore, the GCC must provide its citizens with information geared towards raising awareness and fostering acceptance of the single currency.

The introduction of the single currency will have many economic benefits. The single currency is expected to decrease consumer prices.[88] Business establishments will have lower overhead costs because they will no longer pay the exchange commission charged when dealing with establishments in other countries. Lower overhead costs will produce lower prices for consumers, resulting in an increase in purchasing power for all doing business with member states.[89] More purchasing power may lead to economic growth. An added benefit resulting from the introduction of the single currency is that the commercial market may become more competitive.[90] With a single currency, the price differentials in goods among the GCC member states will disappear. Consumers will benefit because they will be better able to compare the prices of goods. The single currency will eliminate the practice of pricing goods in foreign currencies, and help consumers compare product prices.

---

[87] Currency goes to the heart of people's lives. Currency relates to security, retirement, and kids education. The general population considers the right to issue currency is one of the attributes of an independent country. Assenting to the introduction of the single currency can be regarded as assenting to a partial loss of identity and a surrender of sovereignty. See Robert A. Mundell, Monetary Unions and the Problem of Sovereignty, 579 Annals 123, 134 (2002).
[88] See The Petrodollar Peg; Economics Focus, the Economist (Dec. 9, 2006).
[89] See Erbas G. Abed and Guerami, B. S., The GCC Monetary Union: Some considerations for the Exchange Rate Regime, IMF Working Paper No 66 (2003).

Furthermore, a single currency among GCC countries could reduce or eliminate price fluctuations and will encourage companies to plan their trading and long-term investments.[91] Companies will be able to rely on the single currency and not fear that trade agreements will become unprofitable because of huge price fluctuations. Thus, contracts will become more stable. A stronger GCC economy and improved trading with the single currency could also result in rapid growth in the financial centers of the GCC.[92] Interest in GCC securities could grow because the single currency carries less "currency risk" than the currencies of the individual member states. With less currency risk, investors have a more stable view of a GCC company's credit and are more likely to trust an investment in that company.

Although the introduction of the single currency could be economically beneficially, there may be problems as well. The implementation of the single currency may lead smaller and less efficient companies out of business. Larger competitors may have to cut the labor force and streamline their operations to stay competitive with their new rivals from other GCC countries.[93] There could be also problems with regard to the continuity of contracts. The GCC must adopt measures to assure that contracts will remain unaffected by the introduction of the single currency.[94] The purpose of the principle of continuity of contracts is to bar any claim for rescission, cancellation or non-performance under national law based on the law of frustration (*force majeure* or *Quwa Qahira*), impossibility, or inequity.[95] Additionally, the single currency system of the GCC may create technological problems. Computer systems in the GCC may not be prepared to convert the national currencies into the single currency. Problems also could arise with automatic teller machines (ATM), as banks will be changing currencies into the single currency.

---

[90] See E. Jadresic, On a Currency for the GCC Countries, IMF Policy Discussion Paper (PDP/02/12) (2002).
[91] *Id*.
[92] *Id*.
[93] See Beverly Carl and Roberto MacLean, Jr., Trade and Development, 12 Law & Bus. Rev. Am. 473, 478 (2006).

When the single currency is implemented, member states will produce millions of notes and coins as the new legal tender. The design of these new notes and coins should not be underrated.[96] The single currency coin could have a common GCC face on one side and a national motif of the member state's own choice on the other. The purpose of allowing national motifs on the coins is to retain some link with the past. Also other notes or coins could depict the GCC without national borders, a symbol which is meant to foreshadow the future of the GCC.

Now that the framework of the monetary union have been described, it is important to discuss the relationship between the member states opting out of the monetary union and the GCC as a whole.[97] Two member states -Oman and UAE- have opted out of the early stages of the monetary system.[98] For those member states, there truly is a GCC *a la carte* because they have chosen to adopt the elements which best suited them. The underlying reason for opting out could be that the UAE and Oman do not want GCC integration in the future to lead to a federal form of government and thus to a close form of cooperation.[99] It has to be noted that the GCC Charter does not foresee the possibility for later withdrawal rather it is something that could only have been negotiated beforehand.[100] This may perhaps leave the possibility for future "new" member states to negotiate an opting-out before they become a member state.

---

[94] For example the GCC could state that the introduction of the single currency will not have the effect of altering any term of a legal instrument or of discharging or excusing performance under any legal instrument, nor give a party the right unilaterally to alter or terminate such instrument.
[95] See S. E. Rayner, The Theory of Contracts in Islamic Law: A Comparative Analysis with Particular Reference to the Modern Legislation in Kuwait, Bahrain, and the United Arab Emirates 146, 254, 259-263 (1991).
[96] See George M. Von Furstenberg, One Region, One Money? 579 Annals 106, 113 (2002).
[97] An important aspect of concerning the monetary union is the flexibility granted to GCC member states. The best examples can be found in the ability of member states to opt out of the monetary union.

As far as the relationships between the member states opting out and the GCC is concerned, it is obvious that the UAE and Oman do not participate in all decision-making because they are left out of the decision-making process concerning the monetary union. In addition this fact plays a part in practice because they have shown they are not willing to participate in a further integration process at this time. It is inevitable that such a fact influences GCC relationships and politics generally.

At any rate, in creating the single currency, the GCC has decided to occupy a place on the international scene that is commensurate with their history and their economic strength. In so doing, the GCC demonstrates its unity to the rest of the world and confirms its presence in the monetary sphere. The GCC can play an important role in international institutions responsible for the international monetary system.

---

[98] In 2009, the UAE announced it was pulling out of talks for monetary union. Oman had already withdrawn from the monetary union project in 2007. See Disunited Arab Emirates; Monetary union in the Gulf, *supra* note 48.
[99] The UAE and Oman gave no reason for the decision to withdraw from the monetary union. *Id*.
[100] See the Cooperation Council Charter, *supra* note 10.

**IV. The Relationship between GCC and the WTO**

The GCC does not exist in legal vacuum. Rather, GCC is part of the wider corpus of GATT/WTO law.[101] However, it was not until recently that GCC was notified to the WTO. In November 2006, Saudi Arabia notified the GCC to the WTO.[102] The GCC was notified as GATT article XXIV customs union.[103] Since the GCC was lately notified, there are no reviews yet on the agreement, submissions or comments from other WTO members.

In 2008, the GCC voided GATT article XXIV notification.[104] At the same time, the GCC notified the Committee on Trade and Development (CTD) under the Enabling Clause.[105] Now, the GCC will be reviewed under the Enabling Clause. After the change in notification made by the GCC, other WTO members seemed bewildered and wary. For example, the EU requested for further elaboration on the reasons for this change in notification.[106] The GCC could have determined that its customs union qualifies under GATT article XXIV and the Enabling Clause, but chose the Enabling Clause given the developing country status of GCC members. In the alternative, the GCC could have thought that its customs union would not pass under Article XXIV, but would under the Enabling Clause. Therefore, the GCC changed its notification.

---

[101] The WTO Appellate Body in the United States-Reformulated Gasoline case stated regarding article 3.2 of the Dispute Settlement Understanding that "direction reflects a measure of recognition that the General Agreement on Tariff and Trade is not to be read in "clinical isolation" from public international law." See Appellate Body Report, United States-Standards for Reformulated and Conventional Gasoline, April 29, 1996, WTO Doc. No. WT/DS2/AB/R, at 17.
[102] See WTO Committee on Regional Trade Agreements, Gulf Cooperation Council Customs Union – Notification from Saudi Arabia, November 20, 2006, WTO Doc. No. WT/REG222/N/1.
[103] *Id*.
[104] See WTO Committee on Regional Trade Agreements, Gulf Cooperation Council Customs Union – Notification from Saudi Arabia – Corrigendum, March 31, 2008, WTO Doc. No. WT/REG222/N/1/Corr.1. See also Ben Sharp, Comparing Preferential Trade Agreement Scrutiny under GATT Article XXIV and the Enabling Clause: Lessons Learned from the Gulf Cooperation Council, 7 Manchester Journal of International Economic Law 56, 57 (2010).
[105] See WTO Committee on Trade and Development, Notification of Regional Trade Agreement, March 31, 2008, WTO Doc. No. WT/COMTD/N/25.
[106] See WTO Committee on Trade and Development, Gulf Cooperation Council Customs Union – Saudi Arabia's Notification (WT/COMTD/N/25): Communication from Saudi Arabia, July 18, 2008, WTO Doc. WT/COMTD/66.

To confuse matters further, after the GCC withdrew its customs union notification under GATT article XXIV and notified it under the Enabling Clause, it re-notified the GCC customs union again under GATT article XXIV.[107] The criticism directed at the GCC could have caused it to return to the original notification under article XXIV.

There are different scenarios that would result if the GCC is notified under article XXIV of GATT or under the Enabling Clause because of the differences between these two systems. GATT article XXIV condoned the establishment of free trade areas or customs union subject to several stringent conditions. For example, any agreement must include a plan and schedule for the formation of a free trade area or customs union and the formation should be achieved within a "reasonable length of time."[108] The issue of "reasonable time" was directly addressed during the Uruguay Round negotiations, where it was decided that ten years was a reasonable length of time.[109] The GCC led to the establishment of free trade area among its members in 1983 and customs union in 2003.

Article XXIV of GATT requires any contracting party deciding to enter into a free trade area, or an interim agreement leading to the formation of such an area, to promptly notify the GATT/WTO. This procedural requirement is intended to ensure the transparency of the proposed agreements to other WTO members and provide any necessary information for the examination of the agreements under article XXIV by the Committee on Regional Trade Agreements. The practice in terms of the timing of notification has varied.[110] With regard to the GCC, it should have been notified to the WTO some time ago. The GCC, however, was only notified to the WTO in November 2006 despite the fact that it entered into force in 2003.

---

[107] See WTO Committee on Regional Trade Agreements, Gulf Cooperation Council Customs Union – Notification from Saudi Arabia – Revision, November 17, 2009, WTO Doc. No. WT/REG276/N/1/Rev.1.
[108] The word "reasonable," has caused much confusion in its interpretation. There was no agreement on just how much time was reasonable. For instance, the Greece-EEC Associations Agreement provided for an interim period of twenty-two years before final formation. See Association of Greece with the European Economic Community, Nov. 15, 1962, GATT B.I.S.D (11th Supp.) at 149-50 (1963). Sharp, *supra* note 104, at 59.
[109] See Understanding on the Interpretation of Article XXIV of the General Agreement on Tariffs and Trade 1994.

According to the drafters of GATT article XXIV, the objective of trade regionalism lies in complementing the global trading system. That is, regional free trade agreements and customs unions are to increase trade, not raise barriers to trade with third countries. Moreover, GATT article XXIV requires the free trade area or customs union to eliminate trade barriers on "substantially all" trade among members.[111] Because the GCC was notified under article XXIV, the WTO Committee on Regional Trade Agreements will examine and scrutinize this agreement more extensively to ensure that the GCC does not adversely affect the interests of non-members and to determine how much trade diversion it created, if any.[112]

As a general rule, article XXIV applies only to members of the WTO. For example, the notification requirements of article XXIV apply to the WTO members of GCC, but not to non-WTO members.[113] Preferential agreements with non-members are treated under article XXIV.10 of GATT. Even if article XXIV is considered to be applicable with regard to the GCC, there is also the possibility of a waiver under article XXIV.10 of GATT. Paragraph 10 states that proposals for free trade areas not meeting the criteria described in paragraphs 5 to 9 of article XXIV may be approved by a two-thirds majority of the contracting parties, provided that such proposals eventually lead to the formation of a free trade area. The drafting history indicates that paragraph 10 of article XXIV was intended to provide for the supervision free trade areas in which not all participants were GATT contracting parties.[114] Moreover, it had been shown in practice that the concept "territories of contracting parties" included in article XXIV.5 of GATT had not been interpreted as restricting the ability of establishing free trade areas which include non-GATT members.[115] In most respects, a free trade agreement that complies with article XXIV for WTO members would likely comply for the other free trade agreement members as well since it may be difficult to envision free trade agreement provisions that are different for WTO members from the other non-WTO members.

---

[110] The Treaty of Rome was signed on March 25, 1957 and notified to the Contracting Parties immediately thereafter, with the Treaty entering into force on January 1, 1958. See WTO Secretariat, Regionalism and the World Trading System 12-13 (1995).

[111] Discussions in GATT Working Parties have centered on whether the concept of "substantial' should be understood in qualitative terms (no exclusion of major sectors) or in quantitative terms (percentage of trade of the members covered). See World Trade Organization, Analytical Index: Guide to GATT law and Practice, Vol. 2, 824-27 (1995).

[112] On February 6, 1996, the WTO General Council decided to establish the Committee on Regional Trade Agreements. Under its terms of reference, the Committee on Regional Trade Agreements is mandated to examine regional trade agreements referred to it by the Council for Trade in Goods. See Committee on Regional Trade Agreements - Decision of 6 February 1996, WTO Document No. WT/L/127, paragraph 1.a (February 7, 1996).

Article XXIV.12 of GATT secures the observance of its trade rules by regional and local government authorities. The WTO members of the GCC must ensure that GATT principles are observed by regional and local governments within the territories of those GCC members who are also WTO members. No WTO member is responsible under article XXIV.12 for regional and local governments that are not within that WTO member's territory.

---

[113] Saudi Arabia was latest Gulf country to join the WTO. See Gary G. Yerkey, USTR Announces Bilateral Agreement Clearing Way for Saudi Arabia to Join WTO, 22 International Trade Reporter 1481 (September 15, 2005). Members of the GCC who are also members of the WTO are: Bahrain, Kuwait, Oman, Qatar, Saudi Arabia, and UAE. See WTO, Members and Observers, available at <http://www.wto.org/english/thewto_e/whatis_e/tif_e/org6_e.htm> (last visited January 3, 2010).
[114] See WTO Secretariat, *supra* note 110, at 10.
[115] For example, France obtained a waiver in March 1948 for its proposed customs union with Italy, which was not a contracting party to the GATT at that time. In another example, the Working Party on EEC-Agreements of Association with Tunisia and Morocco approved the established the free trade area although Morocco had no relation yet to the GATT at the time. See World Trade Organization, *supra* note 111, Vol. 2 at 798-799, 829.

Article XXIV is not the only GATT rule that permits the formation of regional trade agreements and customs unions.[116] The Enabling Clause, agreed to during the Tokyo Round, provides for the formation of regional trade agreements among developing countries. There is no difference in the notification required by free trade agreements or customs unions under GATT article XXIV and the Enabling Clause.[117] However, a plan and schedule for implementation of the free trade agreement or customs union is required under article XXIV:5(c), but not under the Enabling Clause. The Enabling Clause includes more lenient criteria compared with GATT article XXIV. For example, unlike article XXIV of GATT, the Enabling Clause drops the conditions on the substantial coverage of trade and allows developing countries to reduce tariffs on mutual trade in any way they wish.[118] The lack of the substantially all trade requirement provides flexibility for members in determining the pace of tariff reduction and product coverage.

Since all members of the GCC are developing countries, it would be covered by Enabling Clause, paragraph 2(c) which permits a regional agreement that do not meet the requirements of GATT article XXIV. If the GCC is examined under the Enabling Clause, the agreement would fall within the jurisdiction of the Committee on Trade and Development. On the other hand, if the GCC is examined under GATT article XXIV the WTO Committee on Regional Trade Agreements would evaluate the GCC customs union.[119] Because the GCC is notified under both GATT article XXIV and the Enabling Clause and the latter notification has not been withdrawn yet, it could lead to the possibility that the GCC customs union could be evaluated under article XXIV and the Enabling Clause. No precedent exists regarding dual examination.[120] In addition, dual examination should be avoided to eliminate the possibility of conflicting rulings. If, for example, the Committee on Trade and Development and WTO Committee on Regional Trade Agreements came to opposite or different conclusions, this complicated the task of multilateral check on regional trade agreements and customs union. At any rate, time will tell how the GCC notifications and examinations will be handled.

---

[116] See Zakir Hafez, Weak Discipline: GATT Article XXIV and the Emerging WTO Jurisprudence on RTAs, 79 North Dakota Law Review 879, 886, 900-902 (2003).
[117] See Different and More Favorable Treatment Reciprocity and Fuller Participation of Developing Countries, L/4903, art. 4.a (Nov. 28, 1979).
[118] *Id*. art. 2.c.

As stated earlier, the GCC and the WTO are not independent of each other; they are highly interdependent. The GCC interacts with regional and international systems.[121] The GCC integration agreements have paid little attention to its relationships with the WTO.[122] There are references to regional blocks and international organizations and the need for the GCC member states to co-ordinate with each other.[123] However, these references are not enough. These references do not provide an ordered legal framework for the relations between the GCC and WTO. Issues such as the status of WTO law within the GCC, how the multiple commitments of GCC member states under GCC law and WTO law can be reconciled, and the rules for resolving conflicts between WTO law and the GCC law have not been addressed by GCC legal documents. These are important issues for the stability of the GCC's economic integration initiatives. Irreconcilable differences between GCC law and WTO law can be susceptible to challenges under the WTO dispute settlement system.[124] It remains to be seen whether GCC's acts will be susceptible for WTO challenge as the GCC member states progress and strengthen their integration processes.

---

[119] The WTO Committee on Regional Trade Agreements is often criticized for the toothless scrutiny that is often employed on Article XXIV agreements. In fact, there has only been one agreement – between Czechoslovakia and Slovakia in 1992 – that has been judged to be fully compliant with this Article. Many issues, like inadequate staffing or timid questioning by members perhaps attempting to set a low threshold for future scrutiny of trade agreements it is party to, have hampered the work of this Committee, especially when considering the rapid increase in these agreements. See Colin Picker, Regional Trade Agreements v. the WTO: A Proposal for Reform of Article XXIV to Counter this Institutional Threat, 26 University of Pennsylvania Journal of International Economic Law 267, 278 (2005).
[120] See Sharp, *supra* note 104, at 58.
[121] See Joost Pauwelyn, Overlaps with the WTO and other Jurisdictions, 13 Minnesota Journal of Global Trade 231 (2004).
[122] There are no statements in the GCC Charter or the Economic Agreement which indicates that member states are conscious of their obligations, as contracting parties to the Marrakesh Agreement Establishing the World Trade Organization.
[123] See Economic Agreement, *supra* note 59, at art. 2.

**V. Relation between GCC and Neighboring Countries**

Since the beginning of GCC's existence, a number of countries have expressed an interest in joining the GCC. Yemen, Iraq, Jordan, and Syria are candidates for inclusion in the GCC.[125] Several reasons seem to make sense for these countries to join the GCC including the cultural similarity between them. All these countries and the GCC share a common language, a generally liberal economic system, and civil law.[126] There is also a great deal of trade between these countries and the GCC.[127] For example, the GCC is the biggest single investor in Jordan.[128] Moreover, large and growing numbers of Arab workers are recruited in the GCC to satisfy employment demands.[129]

---

[124] There are several instances in which regional blocks have been dragged before the WTO Dispute Settlement Body for violating WTO law. See for example Turkey-Restriction on Imports of Textiles and Clothing Products, WTO Doc. No. WT/BS34/AB/R (1999). See also Brazil-Measures Affecting the Imports of Retreaded Tyres, WTO Doc. No. WT/DS332/AB/R (2007).

[125] See Haya M. Awad, Jordan's Accession to the Gulf Cooperation Council: An Analytical Economic Study 2 (2008) (Unpublished master thesis, University of Jordan). See A.F. Ghoneim, Preparing Yemen for Better Economic Integration into the GCC, Ministry of Planning and International Cooperation 13 (2006), available at
<http://www.yemencg.org/library/en/Preparing_Yemen_for_better_economic_integration_into_GCC.pdf>.
(The GCC summit in Saudi Arabia in 2006 announced the possibility of accepting Yemen as a full member in ten years period). The GCC has concern regards Yemen's economic viability.
[126] See John H. Donboli and Farnaz Kashefi, Doing Business in the Middle East: A Primer for U.S. Companies, 38 Cornell International Law Journal 413 (2005). See also Lisa Middlekauff, To Capitalize on a Burgeoning Market? Issues to Consider before Doing Business in the Middle east, 7 Richmond Journal of Global Law & Business 159 (2008).
[127] For instance, Jordan's merchandise trade with GCC amounts to 69.8 of total trade. See Awad, *supra* note125, at 131.

The interests expressed by Jordan and Yemen, among others, raise questions about the GCC accession process itself. The GCC Charter and other documents lack express and detailed provision(s) regarding accession of new members.[130] Also, there is no requirement of geographical nexus between the GCC and new members. Thus, theoretically, Egypt, Lebanon, and Sudan could accede to the GCC. While GCC membership is technically open to others, particularly within the Middle East, the GCC could be reluctant to proceed with accepting new members from other continents.

The expansion of the GCC could prove beneficial. The GCC would form part of an area covering a vast portion of the Middle East and Asian continent. Free access to these larger markets for materials and manufactured products would be highly advantageous.[131] Elimination of tariffs between the GCC and new members would witness an increase trade between the countries concerned. Elimination of other kinds of barriers also would assist in increasing the volume of trade. These trade barriers include different governmental regulations and different health and safety rules.

However, there are could be problems. For example, the GCC has not yet completed the integration process among its own members. Extending the integration process before a template "integration model" has been developed and proven successful to a number of other countries could be a problem with consequences. Engaging in a process of economic integration with Yemen, Jordan, and Syria with different economic and political systems can be challenging.[132] The GCC could create a development fund or bank to aid countries with the most need. The development fund or bank would involve all GCC member states by contributing according to the proportion of each state's economy. This fund or bank would be similar to the fund created for European development that helped raise living standards for Ireland, Spain, and Portugal upon joining the European Union.[133]

---

[128] The largest sources of investment in Jordan come from Saudi Arabia, Kuwait, and the UAE. *Id*. at 70.
[129] See A. Kapiszewski, Arab Versus Asian Migrant Workers in the GCC Countries, UN/POP/EGM/2006/02, Available at <http://www.un.org/esa/population/meetings/EGM_Ittmig_Arab/p02_Kapiszewski.pdf> (2006)**.** About 85 percent of remittances received by Jordan are of GCC origin while in the case of Egypt and Lebanon is about 45 percent and in Syria it reaches 65 percent. See Awad, *supra* note 125, at 75.
[130] There is no provision that would read that any country may accede to the GCC subject to such terms and conditions as may be agreed between such country and the GCC.
[131] See S. Chami, S. Elekdag, and I. Tchakarov, What are the Potential Economics Benefits of Enlarging the Gulf Cooperation Council? IMF Working Paper No. 04/152 (2004) (enlarging the GGC would lead to larger market, lower entry costs, and better market structures).

In addition, cross-border mobility in GCC labor markets could flood local labor markets and lead to racial and ethnic tensions within GCC member states.[134] Further, there is a substantial logistical process involved in implementing the integration process. The integration process covers a large number of different technical areas, economic sectors, and concepts, and involves the involvement of dozens of officials and experts. Supervising the implementation of the GCC common market and negotiating with four different countries at the same time presents a difficult duty for GCC's staff.

---

[132] GCC countries have larger and more prosperous markets than other countries. In addition, the average worker in the GCC is more likely to earn more and pay fewer taxes than the average worker other countries.

[133] In the EU, there are the European Regional Development Fund (ERDF) and the European Social Fund (ESF), collectively referred to as the EU Structural Funds. The level of development funding provided by the EU is impressive. For the period 2007 to 2013, the total funding from EU resources is projected to amount to 347.41 billion Euros (approximately $ 538.5 billion), equal to 35.7% of the total EU budget. Allocations are made to poorer regions wherever they are located in EU Member States. The bulk of ERDF and ESF funding is made available for regions where per capita GDP is below seventy-five percent of the Community average. See Jacqueline Brine, The European Social Fund and the EU: Flexibility, Growth, Stability 11 (2002). See also EU Directorate General for Regional Policy, Funds Available, <http://ec.europa.**eu/** regional policy/policy/funds/index en.htm> (last visited Aug. 20, 2010).

[134] Wealthier GCC member states could fear that other Arab workers would flood their labor markets in search of employment and more generous social welfare rights. However, the number Arab workers living in other GCC member states could decline if there is an economic progress in the home countries as a result of integrating with the GCC.

There can be another way to consider the relation between the GCC and its neighboring countries. Perhaps the relation between the GCC and its potential neighboring countries can be thought of as creating an informal long-term relationship, rather than an actual short-term formal relationship. This kind of relationship creates a link and a channel of communication. It could be seen as statements of intent, meant to create, nurture, and establish a generalized relationship and a tradition of cooperation among these countries, rather than an actual free trade area agreement or customs union within the short term. In this case, the formal relationship can follow after the GCC has developed its own integration template for expansion.

The bottom-line question is whether the GCC is developed and successful enough to be able to expand and absorb other economies can be extremely hard to answer. It is a question with which the EU has been struggling for many years.[135] Perhaps the only way in which it can be answered is retrospectively; if the GCC can successfully negotiate free trade agreements with other countries and if these agreements are successful, then it was developed and successful enough to expand.

The GCC needs to examine its potential membership negotiations with interested countries to determine whether any of its norms, negotiating styles, or positions, or other factors, have negatively affected its ability to implement agreements or successfully conclude membership negotiations with potential new members. Also, the GCC must develop the necessary mechanisms and funds to narrow the gap between existing GCC member states and newly acceded countries. The GCC must find ways to order the relations among native and Arab populations within the GCC. Additionally, the GCC needs to arrive at a concrete and clear framework on expansion. As things stand today, there are vagueness of membership terms and absence of criteria for GCC accession.

---

[135] See Carson W. Clements, More Perfect Union? Eastern Enlargement and the Institutional Challenges of the Czech Republic's Accession to the European Union, 29 Syracuse Journal of International Law & Commerce 401 (2002). See also Peter Katz, The Treaty of Nice and European Union Enlargement: The Political, Economic, and Social Consequences of Ratifying the Treaty of Nice, 24 University of Pennsylvania Journal of International Economic Law 225 (2003).

**VI. Recent Developments**

The U.S has put an increased emphasis on creating free trade areas with GCC member states. The U.S. Administration proposed the establishment of a U.S.-Middle East free trade area by 2013.[136] The U.S. has also signed bilateral trade agreements with Bahrain (2004) and Oman (2006).[137] There can be several benefits accruing through such bilateral trade agreements which include: enhancing goods and services trade; stimulating investment flows;[138] extending standards on intellectual property rights, labor, and the environment; and addressing geopolitical concerns.

The decision by Bahrain and Oman to sign free trade agreements with the U.S. caused a rift with other GCC member states.[139] These free trade agreements could be used by the U.S. to gain preferential market access to the GCC. For example, U.S. exporters could use a bilateral free trade agreement with Bahrain or Oman to gain duty free access to the UAE market.

Moreover, the U.S. trade agreements with GCC member states may create bilateral trade pattern (trade diversion) which discourages intra-GCC economic ties.[140] If all GCC

---

[136] See Yerkey, *supra* note 8.
[137] See Rossella Brevetti, House Approves by Large Margin U.S.-Bahrain Free Trade Agreement, 22 Int'l Trade Rep. (BNA) 2031 (Dec. 15, 2005). See Gary G. Yerkey, United States Announces Completion of Free Trade Negotiations with Oman, 22 Int'l Trade Rep. (BNA) 1603 (Oct. 6, 2005).
[138] See Gary G. Yerkey, U.S. and Bahrain Sign Agreement to Spur Bilateral Trade and Investment, 19 Int'l Trade Rep. (BNA) 1105 (June 20, 2002).
[139] See Gary G. Yerkey, U.S. will Continue to Support GCC while Negotiating Bilateral Free Trade Pacts, 22 Int'l Trade Rep. (BNA) 197 (Feb. 3, 2005).
[140] Economists analyze trade liberalization by considering both trade creation and trade diversion. Trade creation occurs when lower-cost imports from one trading partner replace domestic production from the other. Trade diversion occurs when lower-cost imports from a third party are prevented from entering a

countries do not have comparable free trade agreements with each other, i.e. if they do not conclude a single free trade area, then the common denominator will be the U.S. In a system of hub-and-spokes, trade between each spoke and the hub will be more than trade among the spokes themselves. Therefore, the hub-and-spokes issue has the potential to dramatically reinforce and expand U.S. influence.

An issue could be raised with regard to GCC member states negotiating individually free trade agreements with other countries and whether the GCC, as an institution, has legal authority to prevent these free trade agreements. There are indications that the GCC would have the ability to block such agreements. The GCC adheres to a common commercial policy that is binding on member states.[141] This common commercial policy is based on principles intended to harmonize economic and market interests among GCC member states. If GCC member states had the ability to negotiate free trade agreements on their own, this would frustrate the purpose of the common commercial policy. The common policy also authorizes the Supreme Council to "approve the bases" for dealing with other states or international organizations.[142] Furthermore, the GCC Economic Agreement requires member states to collectively conclude economic agreements with trading partners.[143] This indicates that GCC member states are not allowed to enact agreements that would endanger established GCC law.

---

signatory country because of tariffs or non-tariff barriers. See Michael J. Trebilcock & Robert Howse, The Regulation of International Trade 130 (London: Routledge 1999).
[141] See Economic Agreement, *supra* note 59, at art.1.

[142] See the Cooperation Council Charter, *supra* note 10, at art. 8.
[143] See Economic Agreement, *supra* note 59, at art.2.

As a subject of international law, countries are free to enter into relations and agreements with other countries.[144] Thus, GCC member states have to enjoy treaty-making power. Any act to the contrary by the GCC would infringe upon an area that is considered fundamental to member states' existence. In addition, the WTO encourages member states to enter into free trade agreements.[145] As a party to the WTO, Saudi Arabia would seem to be violating WTO rules if it prevented Bahrain or Oman from entering into trade agreements. However, Bahrain and Oman are members of the GCC, which means that they are bound to conform their interests to the GCC common commercial policy. If Saudi Arabia, for example, considers that Bahrain and Oman free trade agreements are against the common commercial policy and prevent them from concluding additional trade agreement, Bahrain or Oman could initiate a dispute settlement with the WTO. Although the WTO encourages free trade agreements, it is not clear how the WTO would resolve the dispute. It is not likely that the WTO would intervene in an internal dispute.

The GCC should keep close eye on bilateral deals between its member states and outsiders. To achieve this goal, clear rules must be developed so as to determine when the GCC has exclusive power to negotiate on behalf of all its member states and when member states can negotiate international agreements jointly with the GCC to maintain their sovereignty. The legal consequences for seeking and concluding free trade agreements by GCC member states individually should not be underestimated.

---

[144] See Montevideo Convention on Rights and Duties of States, December 26, 1933, art. 1, 49 Stat. 3099, 3100 (The State as a person of international law should possess the following qualifications: (a) a permanent population; (b) a defined territory; (c) government; and (d) capacity to enter into relations with other States).

[145] The GATT provides states should not be refrained from entering into free trade agreements. See Understanding on the Interpretation of Article XXIV of the General Agreement on Tariffs and Trade 1994, *supra* note 109.

## VII. Conclusion

Arab countries have embarked upon ambitious continental integration efforts designed to fulfill their developmental goals. The formation of regional trade agreements in Arab countries and with the GCC was driven by the same economic, political, and security considerations. However, by and large, the GCC is generally regarded as a success insofar as its incarnations as a free trade area, a customs union, and a common market are concerned. The growth of the GCC has been measured and gradual. GCC member states have experienced both positive and negative aspects associated with their integration agreements. Overall, trade among GCC member states increased, strengthening the economic ties among them. However, as a monetary union with a single currency, the verdict remains out.

The quality of the GCC governance structure is most important. The Supreme Council is the most powerful GCC institution and is the head of the GCC governance structure. The authority of the Supreme Council is a general one but not is very detailed such as in the case of conflict management and other emergencies. The GCC Secretariat appears to have no substantive decision-making power and works as an intermediary to facilitate state-to-state discussions, negotiations and monitoring. The Secretariat role can be viewed as one of servicing negotiations and oversight of obligations. In order to be effective, GCC rules must be interpreted in a consistent manner and must be effectively enforced. Unfortunately, GCC document simply do not have an effective dispute-resolution mechanism. The degree of specificity and detail is wanting. There is a need for a panel of permanently appointed, objective, professional panelists ensuring a consistent interpretation of the GCC's legal norms. This perception ensures that its rulings will be followed by all member states. The absence of an effective and trustworthy dispute-resolution and rule enforcement mechanism creates major obstacles to the implementation and expansion of the GCC.

Fundamental to the formation of an integrated GCC is the creation of a common legal system. Harmonization of laws facilitates transactions concluded between different sovereignties and encourages all forms of economic exchange. Having taken this into consideration, and in order to achieve its objectives and aims of establishing a regional block, the GCC attempted to harmonize various laws and regulations.
The introduction of the single currency will affect every aspect of GCC structure and life. Therefore, there must be an interim period that serves as a period designed to ensure a smooth changeover from national currencies to the single currency. The general population should become familiarized with the single currency during the transitional period. During this stage, it is important to ensure that the GCC Central Bank will have total independence in its decision-making. The GCC is characterized by the ability to selectively exit from obligations in specific circumstances considered to be inconsistent with national interests. If Oman and UAE felt uncomfortable with the arrangement within the GCC they could opt out.

The GCC faces many challenges ahead. The GCC does not exist in legal vacuum. Rather, GCC is part of the wider corpus of GATT/WTO law. The GCC integration agreements have paid little attention to its relationships with the WTO. Issues such as the status of WTO law within the GCC, how the multiple commitments of GCC member states under GCC law and WTO law can be reconciled, and the rules for resolving conflicts between WTO law and the GCC law have not been addressed by GCC legal documents.

Relation between GCC and neighboring countries is another sticky issue. The GCC Charter and other documents lack express and detailed provisions regarding accession of new members. There is no requirement of geographical nexus between the GCC and new members. Furthermore, the GCC has not yet completed the integration process among its own members. Extending the integration process before a template "integration model" has been developed and proven successful to a number of other countries could be a problem with consequences.

The U.S. signed bilateral free trade agreements with some GCC member states, namely Bahrain and Oman. The decision by Bahrain and Oman to sign free trade agreements with the U.S. caused tensions with other GCC member states. These free trade agreements could be used by the U.S. to gain preferential market access to the GCC. An issue could be raised with regard to GCC member states negotiating individually free trade agreements with other countries and whether the GCC, as an institution, has legal authority to prevent these free trade agreements. By the same token, as subjects of international law, Bahrain and Oman are free to enter into relations and agreements with other countries. If Saudi Arabia, for example, considers that Bahrain and Oman free trade agreements are against the common commercial policy and prevent them from concluding additional trade agreement, Bahrain or Oman could initiate a dispute settlement with the WTO. Clear rules must be developed so as to determine when the GCC has exclusive power to negotiate on behalf of all its member states and when member states can negotiate international agreements jointly with the GCC to maintain their sovereignty.

Although economic integration among GCC member states is an ambitious step in the right direction, there are gaps and challenges ahead. The best way to address the gaps and challenges that exist in formulating integration processes in the GCC is to start with a clear set of rules and put the necessary mechanisms in place. Integration attempts must also exhibit a high level of commitment in order to deflect dynamics of disintegration that have all too often frustrated meaningful integration in Arab countries. If the GCC can address these issues, it could become an economic powerhouse within Arab countries and even Asia.